# Strengthening and Weakening by Dislocations in Monolayer MoS$_2$

Li Yang[1], Jinjie Liu[1], Yanwen Lin[1], Ke Xu[1], Xuezheng Cao[1], Zhisen Zhang[1,*] and Jianyang Wu[1,2,*]

[1]Department of Physics, Research Institute for Biomimetics and Soft Matter, Jiujiang Research Institute and Fujian Provincial Key Laboratory for Soft Functional Materials Research, Xiamen University, Xiamen 361005, PR China

[2]NTNU Nanomechanical Lab, Norwegian University of Science and Technology (NTNU), Trondheim 7491, Norway

**Abstract:** Dislocations govern the properties of any crystals. Yet, how dislocation of pentagon-heptagon (5|7) in grain boundaries (GBs) affects the mechanical properties of two-dimensional MoS$_2$ crystals remains poorly known. Using atomistic simulations and continuum disclination dipole model, we show that, depending on the tilt angle and 5|7 dislocation arrangement, MoS$_2$ GB strength can be enhanced or reduced with tilt angle. For zigzag-tilt GBs primarily composed of Mo5|7+S5|7 dislocations, GB strength monotonically increases as the square of tilt angle. For armchair-tilt GBs with Mo5|7 or S5|7 dislocations, however, the trend of GB strength breaks down as 5|7 dislocations are non-evenly spaced. Moreover, mechanical failure initiates at the bond shared by 5|7 rings, in contrast to graphene where failure occurs at the bond shared by 6|7 rings. This work provides new insights into mechanical design of synthetic transition metal dichalcogenide crystals via dislocation engineering.

**Keywords:** Molybdenum disulfide, Grain boundary, Tensile strength, Molecular dynamics, Continuum mechanics

*Corresponding authors: zhangzs@xmu.edu.cn, jianyang@xmu.edu.cn

# 1 Introduction

Single-layer molybdenum disulfide (SLMoS$_2$) is a representative two-dimensional crystal of transition metal dichalcogenides (TMDs). SLMoS$_2$ has attracted scientific and technological interests in recent years due to its fascinating electronic[1], optical[2], thermal[3] and mechanical properties[4, 5]. For example, pristine SLMoS$_2$ shows distinctive characteristic of semiconductors with a direct bandgap of about 1.8 eV[6]. It was identified to exhibit high in-plane Young's modulus of $270 \pm 100$ GPa[7], and high in-plane strength comparable to that of stainless steel. As a result, it found a host of important practical applications, from conventional applications, for example, lubricant[8] and catalysis[9], to next-generation device systems, such as field-effect transistors[10], phototransistors[11], optoelectronics [12-14], and mechanically flexible electronics[5, 15-17].

To date, fabrication of large-area single crystal of SLMoS$_2$ by the state-of-art exfoliation and chemical vapor deposition (CVD) growth techniques remains a big challenge. For instance, exfoliated monocrystalline SLMoS$_2$ are commonly flakes with small dimensions and native defects-contained. Because of multiple nucleation sites, CVD-grown samples of SLMoS$_2$ are large-area polycrystalline 2D materials that are structurally characterized by a patchwork of crystalline grains joined by grain boundaries (GBs). However, GBs in SLMoS$_2$ are mainly dominated by a variety of topological lattice defects [2, 18-21], including 4|4, 4|6, 8|4|4, 5|7, 6|8 defective pairs[19-24], resulting in significant change in electronic, thermal and mechanical properties[25-28]. For example, twin GB with 4-4-8 rings is able to show conflicting benefit/detriment in the electronic properties of SLMoS$_2$; At low temperature, mid-gap states are localized, leading to improvement of in-plane electrical conductivity, whereas at high temperature, the carries are scattered around the GB, thereby impeding the transport capacity[27]. As a result of unique arrangement of GBs, polycrystalline MoS$_2$ films are able to exhibit ultralow thermal conductivity of 0.27 W·m$^{-1}$·K$^{-1}$ close to the value as that of Teflon[29].

In mechanics, similar to the case of graphene[30-36], it was identified that in-plane mechanical properties and failure behavior of SLMoS$_2$ are greatly affected by the type of vacancy defects, dislocations, GB and crystalline grain sizes in its planar structure[2, 21, 25, 26, 37-44]. For instance, sulfur (S) vacancy in SLMoS$_2$ clearly degrades the stiffness and strength[37, 41, 42], but enhances the mechanical ductility[38]. With increasing the concentration of point vacancy or molybdenum (Mo) antisite defects from 0 - 25%, elastic properties are monotonically reduced[45]. In contrast, investigations on the effects of dislocations, GB and grain size on the mechanical characteristics of SLMoS$_2$ are very limited. Uniaxial tensile characteristics of SLMoS$_2$ can be greatly dictated by polar dislocation core; SLMoS$_2$ containing different type of dislocation core shows distinct mechanical properties, and SLMoS$_2$ with armchair 4-8 dislocations presents completely brittle fracture, but with zigzag-oriented dislocations exhibits characteristic dual brittle/ductile fractures[25]. Under nanoindentation, SLMoS$_2$ with unique 4|4, 4|8 or 8|8 GB structure was initially failed at the GBs and the breaking force was found to be independent of GB energy[2]. With regard to the case of grain size, it was intriguingly identified that both tensile stiffness and tensile strength of polycrystalline SLMoS$_2$ to the grain size are described by inverse pseudo Hall-Petch relation[38, 43].

Despite the aforementioned advanced investigations, fundamental mechanical characteristics of defects-contained SLMoS$_2$ are largely insufficient. For example, a critical issue remains unexplored as how twin GBs influence the mechanical properties of SLMoS$_2$ that is of both scientific and technological importance for its practical applications in device systems. To address this critical question, SLMoS$_2$ bicrystals containing a set of armchair- and zigzag-oriented twin GBs primarily composed of polar pentagon-heptagon (5|7) pair defects (Figure 1) are constructed and the dependence

of twin GB strengths on grain misorientation angle is systematically examined by classic molecular dynamic (MD) simulations and continuum mechanics analysis.

## 2 Atomic Models and Methodology

As is known, SLMoS$_2$ is a bielemental structure with top-viewed motif similar to single layer boron nitride (SLBN), and is structurally characterized by a sandwiched structure of a monatomic Mo-layer between two monatomic S-layers, in contrast to monoelemental graphene. As proposed by Wei et. al[30], for twin GB structure-dominated graphene, SLMoS$_2$ bicrystals containing twin GB structures are constructed from straightforward "cut-stitch" procedure. Because of hexagonal lattice of MoS$_2$ similar to graphene, there are zigzag- and armchair-tilt twin GB structures composed of 5|7 dislocation array. Moreover, as a result of its bielemental composition, polar 5|7 dislocation can be Mo- and S-oriented, with pentagon and heptagon rings shared by Mo-Mo and S-S bonds, respectively, thereby being marked as Mo5|7 and S5|7[25].

To quantitatively characterize both zigzag- and armchair-tilt twin GB structures of MoS$_2$, tilt angle ($\theta$) of twin GBs is defined. For zigzag-tilt twin GB structures, the tilt angle $\theta$ is determined as

$$\theta = 60° - 2arcsin\frac{|\vec{b}_{(1,0)+(0,1)}|}{2d_{(1,0)+(0,1)}} \tag{1}$$

where $|\vec{b}_{(1,0)+(0,1)}|$ is the width of dislocation, and $d_{(1,1)}$ is the density of dislocation. Here, the length of Burge vector $|\vec{b}_{(1,0)+(0,1)}|$ is equal to 9.11 Å. For armchair-tilt twin GB structures, the tilt angle $\theta$ is expressed as

$$\theta = 2arcsin\frac{|\vec{b}_{(1,0)}|}{2d_{(1,0)}} \tag{2}$$

where $|\vec{b}_{(1,0)}|$ and $d_{(1,0)}$ are the width and density of dislocation, respectively. Here, the length of Burge vector $|\vec{b}_{(1,0)}|$ is 4.23 Å.

Figure 1 shows snapshots of a set of top-viewed twin GBs configurations taken into investigation in this work. Figure 1a presents zigzag-tilt twin GB structures with $\theta$ varying from 40.02 - 54.25°, where twin GB structures are dominated by clusters of polar Mo5|7 + S5|7 dislocations. Whereas, Figure 1b and c display armchair-tilt twin GB structures mainly composed of polar Mo5|7 and S5|7 dislocation arrays with $\theta$ ranging from 5.06 - 21.98°, respectively. Periodic boundary condition (PBC) is applied in the dislocation array direction to mimic infinite straight twin GBs. The edge length of all twin GB-contained SLMoS$_2$ in the other planar direction is constructed to be around 150 nm for eliminating spurious effects of non-PBC.

Prior to MD relaxations, as-constructed twin GBs-contained SLMoS$_2$ are quasi-statically relaxed to a local energy-minimum state through conjugate gradient method with energy and force tolerances of $1.0 \times 10^{-4}$ eV/mol and $1.0 \times 10^{-4}$ eV/(mol·Å), respectively. Afterwards, the samples are globally relaxed by MD runs with 100000 timesteps at temperature of 10 K and zero pressure in the planar directions under isothermal-isobaric ensemble. Initial velocities of Mo and S atoms in the samples are imposed following Gaussian distribution at the given temperature of 10 K. Lastly, uniaxial straining along one planar direction that is perpendicular to 5|7 dislocation array is performed at temperature of 10 K, and the uniaxial straining rate is reasonably set to be $1.0 \times 10^8$/s. Whereas, zero pressure is maintained along the other planar direction under isothermal-isobaric ensemble, ensuring consideration of Poisson effect during the straining.

In the MD runs, the temperature and pressure are controlled by Nosé-Hoover thermostat and barostat techniques with damping times of 100 and 1000 timesteps, respectively. The velocity-Verlet algorithm with a timestep of 0.5 fs is employed to integrate the Newton's motion of SLMoS$_2$ systems. Mo and S atomic stresses in the SLMoS$_2$ systems are calculated based on the definition of virial stress.

Moreover, it is assumed a thickness of 6.5 Å of SLMoS$_2$ for computing Mo and S atomic stresses[43].

All the molecular mechanics (MM) and MD simulations are accomplished using the Large-scale Atomic/Molecular Massively Parallel Simulator (LAMMPS)[46]. To describe the atomic interactions in the twin GB-contained SLMoS$_2$ systems, MoS$_2$-type Reactive Empirical Bond-Order (REBO) forcefield modified by Wang et. al is employed[47]. Such MoS$_2$-type REBO forcefield has been successfully utilized to mimic the mechanical characteristics of various MoS$_2$ nanostructures[4, 25, 43, 48].

## 3 Results

### 3.1 GB Energy in SLMoS$_2$

Figure 2 summaries the GB energy for Mo5|7, S5|7 and Mo5|7+s5|7 dislocation-dominated GBs with $\theta$ from 5.06/42.03 - 21.98°/54.25° for armchair/zigzag interface. Obviously, GB energy greatly varies with $\theta$ and GB types, ranging from around 2.0 - 7.0 eV/nm, comparable to that for one-atom-thick graphene and hexagonal boron nitride ($h$-BN)[32, 49-51]. Moreover, GB energies predicted by MD simulations are in good agreement with previous DFT calculations [21], for example, energies of armchair-tilt GBs with $\theta$ = 9.40° and 21.8° are 3.7 eV/nm and 5.2 eV/nm, respectively, and zigzag-tilt GB with $\theta$ = 48.35° shows GB energy of 4.9 eV/nm. Those indicate the accuracy of MoS$_2$-type REBO forcefield for predicting SLMoS$_2$ GB configurations. It is observed non-equivalence in the GB energy for armchair and zigzag interfaces, resulting from mismatch in the orientation of 5|7 dislocation in the GB structures. Intriguingly, Mo5|7 dislocation-dominated GBs with identical $\theta$ show higher GB energy than S5|7 dislocation-dominated ones, which is primarily attributed to the fact that double Mo atoms in the center of dislocation are coordinatively-unsaturated, as well as their different chemical composition and three dimensional configurations.

For armchair-tilt GBs, GB energy monotonically increases with increasing $\theta$. However, the increase in GB energy becomes less pronounced with increasing $\theta$, originating from stronger interactions of neighboring dislocations in GBs with higher density of dislocation. In contrast, the variation in the GB energy with $\theta$ is inverse because of different definitions of $\theta$ between armchair- and zigzag-tilt GBs. However, it is summarized that GB energy positively correlates with dislocation density of GB. For both armchair- and zigzag-tilt GBs with small density of 5|7 dislocation, SLMoS$_2$ GB energy can be well-described by the Read-Shockley model[52], with expression as follows

$$\gamma = \frac{bE}{8\pi}\theta'\left[1 + ln\frac{b}{r_0} - ln(2\pi\theta')\right] \qquad (3)$$

where $E$, $b$ and $r_0$ are the Young's modulus, Burger vector and the radius of 5|7 dislocation core of SLMoS$_2$, respectively. Note that $\theta' = \theta$ and $\theta' = 60° - \theta$ for armchair- and zigzag-tilt GBs, respectively. By a least-squares fit to the Read-Shockley equation for S5|7-, Mo5|7 and Mo5|7 + S5|7-dominated GBs, $r_0$ is yielded to be 0.077, 0.146 and 0.050 nm, respectively, where $E = 123$ N/m and $b = 3.19$ Å[53]. The yielding $r_0$ of MoS$_2$ are comparable to that of other 2D structures, for example, $r_0 = 1.20$ Å for graphene[32], $r_0 = 0.54 \pm 0.01$ Å and $0.90 \pm 0.01$ Å for zigzag- and armchair-tilt GBs of $h$-BN[49], respectively.

**3.2 Dependence of GB Tensile Strength on Tilt Angle of GBs**

Figure 3a-c shows the resultant tensile stress-strain curves of all twin GB-dominated SLMoS$_2$. All stretching curves are characterized by that loading stresses are increased with increasing uniaxial strain to critical value, followed by a sudden drop of stretching stress to zero. Such sudden drop is indicative of brittle failure for all samples. However, there is difference in the drop characteristics of loading stress between zigzag- and armchair-tilt twin GB-contained SLMoS$_2$. For armchair-tilt twin

GB-dominated SLMoS$_2$, the drop of loading stresses to zero takes place in very limited strain regime, similar to the case of graphene. For zigzag-tilt types of Mo5|7- and S5|7 dislocation-dominated samples, however, the drop of loading stresses to zero occurs within finite strain regime. Those indicate their distinct brittle failure deformation mechanisms.

Figure 3d-f shows the variations in the uniaxial tensile strength of all SLMoS$_2$ bicrystals with $\theta$. For zigzag-tilt twin GB-dominated SLMoS$_2$, tensile strength of GBs monotonically declines from around 11 - 8.7 N/m by 21% as $\theta$ is increased from 42.03 - 54.25°. Such monotonic trend in tensile strength of zigzag-tilt twin GBs of SLMoS$_2$ is similar to the case of 2D graphene[30]. With regard to armchair-tilt types of Mo5|7 and S5|7 dislocation-contained samples, however, tensile strength of twin GBs non-monotonically changes with $\theta$, breaking usual thinking that strength of materials increases monotonically with density of dislocation defects. For example, as $\theta$ of Mo5|7/S5|7 dislocation-dominated GB samples vary from 5.98 - 13.16°/5.06 - 13.16°, the tensile strength of twin GBs is insignificant to vary with tilt angle. Beyond $\theta = 13.16°$, however, both armchair-tilt types of samples show 'flipped' behavior in the tensile strength, namely, as $\theta$ is increased, their tensile strengths are first reduced but then enhanced, where a local minimum value occurs at critical $\theta = 18.82°$.

Such strength enhancement and weakening by 5|7 dislocation defect was also identified in 2D monoelemental graphene[30]. Similar to the case of monoelemental graphene[30, 31], the normal and abnormal variations in tensile strength of twin GBs with $\theta$ are primarily attributed to unique arrangement and density of 5|7 dislocation core in SLMoS$_2$, as illustrated in Figure 1. For example, for zigzag-tilt twin GBs, a disclination cluster of Mo5|7+ S5|7 dislocations locally coexists and is evenly spaced, resulting in its monotonic change in strength of GBs. Whereas, for armchair-tilt GBs, Mo5|7 or S5|7 dislocation is locally isolated, and it is non-evenly or evenly spaced, depending on $\theta$.

Specifically, armchair-tilt twin GBs with $\theta$ = 5.06°, 5.98°, 7.31°, 9.40°, 13.16° and 21.98° are structurally characterized by an array of evenly distributed 5|7 dislocations, while those with other $\theta$ have an array of non-evenly distributed 5|7 dislocations. The two distinct arrangements of 5|7 dislocation could generate different releases in pre-strain of bonds and strain energy by dislocation interaction[30, 31, 36].

### 3.3 Pre-stress Induced by Dislocations in SLMoS$_2$

To provide insights into the unique dependence of tensile strength of twin GBs on $\theta$, distribution of atomic pre-stress in twin GB samples are analyzed. Figure 4 shows the contours of stress $\sigma_{xx}$ in all investigated samples prior to tensile straining. Apparently, as a result of the presence of 5|7 dislocation defect in SLMoS$_2$, all samples are heterogeneously pre-stressed, particularly at twin GB structures. Such heterogenous atomic pre-stress in SLMoS$_2$ is reduced away from 5|7 dislocation cores, with renormalization to decay for far-field distributions of pre-stress. Moreover, dipolar stress field is identified around isolated Mo5|7/S5|7 dislocation and Mo5|7 + S5|7 dislocation cluster in zigzag- and armchair-tilt GBs, where heptagon- and pentagon-oriented regions show positive and negative values of stress $\sigma_{xx}$, respectively. This suggests that heptagon- and pentagon-oriented regions are pre-stressed in tension and contraction, respectively.

For zigzag-tilt GBs, as shown in Figure 4a, concentration of pre-stress $\sigma_{xx}$ at cluster of Mo5|7 + S5|7 dislocations can be observed to monotonically vary with $\theta$ of twin GBs. With rising $\theta$ of GBs, concentration of pre-stress $\sigma_{xx}$ becomes more pronounced. In terms of density of dislocation defect, pre-stress is less concentrated at highly defective GBs of SLMoS$_2$. This is mainly due to the fact that, as a result of even distribution of clusters of Mo5|7 + S5|7 dislocations in the zigzag-tilt GBs, localized pre-stress at GBs is largely co-offset by strong interactions of dense dislocation defects[34]. Such

inverse relation between pre-stress concentration and dislocation density was also identified in monoelemental graphene[36]. This explains the monotonic reduction in tensile strength of armchair-tilt twin GB samples with rising $\theta$.

For armchair-tilt GBs composed of either Mo5|7 or S5|7 dislocations, however, there is difference in variation of concentration of tensile pre-stress $\sigma_{xx}$ with $\theta$, as illustrated in Figure 4b and 4c. Similar to the case of zigzag-tilt GBs, as Mo5|7/S5|7 dislocation is evenly distributed at armchair-tilt GBs, armchair-tilt GBs show negative correlation between maximum pre-stress and $\theta$, thereby resulting in rising tendency in GB strength with increasing tilt angle of GBs. As Mo5|7/S5|7 dislocation is non-evenly distributed at armchair-tilt GBs, however, armchair-tilt GBs show non-uniformly localized pre-stress at Mo5|7/S5|7 dislocations, with global maximum pre-stress at heptagon that has longer distance to its next neighboring heptagon. And the global maximum pre-stress at heptagon is larger than that in armchair-tilt GBs with evenly-arranged 5|7 dislocations and lower $\theta$. As a result, weakening in tensile strength of GB by specific arrangement of 5|7 dislocation can be identified from Figure 3e and 3f. Intriguingly, there is distinct difference in maximum localized pre-stress between Mo5|7 and S|57 dislocations for armchair-tilt GBs. For Mo5|7 dislocation-dominated GBs, maximum pre-stress occurs at double Mo-atoms that are shared by pentagon and heptagon of dislocation, whereas for S5|7 dislocation-dominated GBs, it appears at single Mo atom of heptagon that has longest distance to its nearest pentagon. From atomic point of view, deformation is characterized by changing atomic distances in structures. Therefore, Mo5|7 dislocation-dominated GBs are easier to mechanically deform than S5|7 dislocation-contained GBs. Those explain that S5|7 dislocation contained-GBs are mechanically robust defective structures.

### 3.4 Fracture Mechanisms of SLMoS$_2$ GBs

To further revealing the failure mechanisms of SLMoS$_2$ containing GBs subjected to mechanical load perpendicular to GBs, their structural developments are recorded. As representatives, a set of snapshots of zigzag- and armchair-tilt GBs with $\theta$ = 42.03° and 17.96° are shown in Figure 7, where the color code is on the basis of the value of *von Mises* stress. At zero strain, GBs are highly stress-concentrated. Upon critical strains, failure of both zigzag- and armchair-tilt GBs initiates via dissociation of 5|7 bonds shared by pentagonal and heptagonal rings, in sharp contrast to graphene GBs that initially fail via breakage of 6|7 bonds shared by hexagonal and heptagonal rings[34]. However, because zigzag-tilt GBs is primarily composed of Mo5|7 + S5|7 dislocation clusters, their failure initially occurs by dissociation of S-S bonds in S5|7 dislocation followed by breaking Mo-Mo bonds of Mo5|7 dislocation, indicating that Mo-Mo bond in 5|7 dislocation is mechanically stronger than S-S bond. Subsequently, as uniaxial tension is further imposed, multi-cracks gradually propagate along the GBs accompanied by neighboring fracture of hexagon, resulting in deep drop of loading stress within finite strain regime. For armchair-tilt GBs with evenly-distributed Mo5|7 or S5|7 dislocations, failure occurs via almost simultaneous breakage of 5|7 bonds shared by pentagon and heptagon, whereas for those with non-evenly distributed Mo5|7 or S5|7 dislocations, failure takes place by dissociation of Mo-Mo or S-S bonds at bottom-most dislocation in 5|7 dislocation clusters as a result of non-uniform stress concentration at a cluster of 5|7 dislocations. Soon afterwards, the following rapid failure is dominated by dissociation of the bonds shared by hexagon and heptagon, similar to initial failure of graphene GBs[31, 34]. Finally, cracks propagate rapidly along the GBs, explaining sudden drop of loading stress to zero within small strain regime. In short, both zigzag- and armchair-tilt GBs fail via brittle fracture behavior, independent on $\theta$.

## 4 Discussions

From perspective of topology, a 5|7 defect pair resembles a disclination dipole that is structurally characterized two disclinations of opposite signs, as shown in Figure 6a. To in-depth understand the dependence of tensile strength of both armchair- and zigzag-oriented GBs in SLMoS$_2$ on $\theta$ from continuum theory, the continuum disclination dipole model (CDDM) developed by Wei et. al[30] for 2D graphene is applied. Although 5|7 dislocations in graphene cause out-of-plane buckling, SLMoS$_2$ containing GBs primarily composed of 5|7 dislocations are perfectly flat 2D sheets. This stresses that CDDM can be effectively adopted for predicting GB strength in SLMoS$_2$ without stretching.

For armchair-tilt GBs consisting of an array of evenly-distributed disclination dipoles, the stress component $s_{xx}$ perpendicular to GB at $(0, \Delta)$, namely, the bond shared by the 5|7 pair, can be expressed from the CDDM as follows[30]

$$\frac{S_{xx}}{\sigma_0} = -\frac{2\pi^2 \Delta d}{3 h_d^2} \frac{\theta^2}{\omega^2} \quad (4)$$

with

$$\sigma_0 = \frac{E\omega}{4\pi}$$

where $\Delta$, $\omega$, $h_d$ and $\theta$ are the distance from the center of the disclination dipole to the critical bond of 5|7, the rotation strength of the disclination, characteristic height and tilt angle, respectively. $E$ is the Young's modulus of pristine SLMoS$_2$.

When armchair tilt GBs composed of an array of non-evenly distributed disclination dipoles, however, GBs show different magnitude in the normal stress components at $(0, \Delta)$ of diploes in GBs. For armchair-tilt GB with $\theta = 16.46°$, in which two diploes are continuously connected, the extra normal stresses at the top and bottom diploes ($C_T$ and $C_B$ in Figure 6b), respectively, predicted from

the CDDM as follows

$$\left(\frac{S_{xx}}{\sigma_0}\right)_T = -\frac{2\pi^2 \Delta d}{3L^2} - \frac{2\pi^2 (h_d+\Delta)d}{3L^2} + \ln\left(\frac{h_d+\Delta+d}{h_d+\Delta-d}\right) \tag{5a}$$

$$\left(\frac{S_{xx}}{\sigma_0}\right)_B = -\frac{2\pi^2 \Delta d}{3L^2} + \frac{2\pi^2 (h_d-\Delta)d}{3L^2} - \ln\left(\frac{h_d-\Delta+d}{h_d-\Delta-d}\right) \tag{5b}$$

For armchair-tilt GB with $\theta = 17.96°$ in which three dipoles are continuously arranged, the extra normal stresses produced by a dislocation cluster at the top, middle and bottom ($C_T$, $C_M$ and $C_B$ in Figure 6c) are, respectively, given from the CDDM as follows

$$\left(\frac{S_{xx}}{\sigma_0}\right)_T = -\frac{2\pi^2 (h_d+\Delta)d}{L^2} + \ln\left(\frac{h_d+\Delta+d}{h_d+\Delta-d}\frac{2h_d+\Delta+d}{2h_d+\Delta-d}\right) \tag{6a}$$

$$\left(\frac{S_{xx}}{\sigma_0}\right)_M = -\frac{2\pi^2 \Delta d}{L^2} + \ln\left(\frac{h_d^2-(\Delta+d)^2}{h_d^2-(\Delta-d)^2}\right) \tag{6b}$$

$$\left(\frac{S_{xx}}{\sigma_0}\right)_B = \frac{2\pi^2 (h_d-\Delta)d}{L^2} - \ln\left(\frac{h_d-\Delta+d}{h_d-\Delta-d}\frac{2h_d-\Delta+d}{2h_d-\Delta-d}\right) \tag{6c}$$

Moreover, for armchair-tilt GB with $\theta = 18.82°$ in which four diploes are continuously placed, the extra normal stresses at the top, middle and bottom ($C_T$, $C_{M1}$, $C_{M2}$ and $C_B$ in Figure 6d) are, respectively, predicted from the CDDM as follows

$$\left(\frac{S_{xx}}{\sigma_0}\right)_T = -\frac{4\pi^2 (3h_d+2\Delta)d}{3L^2} + \ln\left(\frac{h_d+\Delta+d}{h_d+\Delta-d}\frac{2h_d+\Delta+d}{2h_d+\Delta-d}\frac{3h_d+\Delta+d}{3h_d+\Delta-d}\right) \tag{7a}$$

$$\left(\frac{S_{xx}}{\sigma_0}\right)_{M1} = -\frac{4\pi^2 (2h_d+\Delta)d}{3L^2} + \ln\left(\frac{(h_d+d)^2-\Delta^2}{(h_d-d)^2-\Delta^2}\frac{2h_d+\Delta+d}{2h_d+\Delta-d}\right) \tag{7b}$$

$$\left(\frac{S_{xx}}{\sigma_0}\right)_{M2} = \frac{4\pi^2 (2h_d-\Delta)d}{3L^2} - \ln\left(\frac{(h_d+d)^2-\Delta^2}{(h_d-d)^2-\Delta^2}\frac{2h_d-\Delta+d}{2h_d-\Delta-d}\right) \tag{7c}$$

$$\left(\frac{S_{xx}}{\sigma_0}\right)_B = \frac{4\pi^2 (3h_d-2\Delta)d}{3L^2} - \ln\left(\frac{h_d-\Delta+d}{h_d-\Delta-d}\frac{2h_d-\Delta+d}{2h_d-\Delta-d}\frac{3h_d-\Delta+d}{3h_d-\Delta-d}\right) \tag{7d}$$

With regard to zigzag tilt GBs in this study, they are structurally characterized by evenly-distributed disclination cluster of M5|7 + S5|7. As a result, the normal stress of zigzag tilt GBs can also be predicted from the CDDM as equation (4).

On the basis of those CDDM equations, it is known that, upon uniaxial tension perpendicular to GBs, 5|7 dislocation at the top-most diploe of a disclination cluster in SLMoS$_2$ GBs is the most mechanically loaded, in good agreement with MD predictions as shown in Figure 4. As a result, tensile strength of SLMoS$_2$ GBs is given by the strength of a disclination dipole without impact by the other defects in the structure, minus the normal stress by all other disclination dipoles, with mathematical expression as $\sigma_y = \sigma_{y0} - s_{xx}$. Because there exist differences in the equivalent Burgers vectors of dislocations and the orientation of the 5|7 shared bond, armchair- and zigzag-tilt GBs have different $\sigma_{y0}$. Moreover, because armchair-tilt GBs are composed of an array of Mo5|7 or S5|7 dislocations, there is difference in the strength of $\sigma_{y0}$ between Mo5|7- and S5|7-dominated armchair-tilt GBs, differing from the case of graphene. Figure 7 shows the theoretical predictions by CDDM and the results by MD simulations, where the parameters utilized for theoretical predictions are listed in Table 1. Apparently, tensile strengths of SLMoS$_2$ GBs with evenly-distributed disclination dipoles by CDDM with equation (4) are matched with those of MD predictions, particularly for zigzag-tilt GBs. As armchair-tilt SLMoS$_2$ GBs are dominated by non-evenly distributed disclination dipoles, for example, armchair-tilt GBs with $\theta = 16.46°$, $17.96°$ and $18.82°$, the GB strengths are not able to be described by equation (4), but are able to be captured by equations (5)-(7), respectively. In a nutshell, all the details in the GB strength-tilt angle curves of SLMoS$_2$ from MD simulations can be explained by CDDM.

## 5 Conclusions

The mechanical properties of symmetrical tilt SLMoS$_2$ GBs primarily consisting of an array of 5|7 dislocations are explored by both MD simulations and continuum mechanics of disclination dipole model. MD simulations reveal that there are unique relationships between GB strength and tilt angle in SLMoS$_2$ in ways not previously identified for MoS$_2$. Zigzag-tilt GBs that are structurally

characterized by Mo5|7+S5|7 dislocations show increase in GB strength as the square of tilt angle. For armchair-tilt GBs composed of Mo5|7 or S5|7 dislocations, GB strength follows the trend as dislocations are evenly spaced in GB, but breaks down when dislocations are non-evenly spaced. The conflicting mechanical enhancement and weakening behaviors in SLMoS$_2$ by 5|7 dislocations can be described by CDDM. This stresses that mechanical properties of SLMoS$_2$ are dictated not only by density of 5|7 dislocation, but also by arrangement of 5|7 dislocation. Moreover, all GB samples fail via tension-induced dissociation of the bond shared by 5|7 rings, not the bond shared by 6|7 rings identified in graphene. This study advances fundamental understanding how dislocation defects interact and influence the mechanical properties for 2D MoS$_2$ crystals.

## Acknowledgments

This work is financially supported by the National Natural Science Foundation of China (Grant Nos. 11772278 and 11904300), the Jiangxi Provincial Outstanding Young Talents Program (Grant No. 20192BCBL23029). Y. Yu and Z. Xu from Information and Network Center of Xiamen University for the help with the high-performance computer.

**Tables and caption**

Table 1 Geometrical and material parameters of both zigzag- and armchair-tilt grain boundaries (GBs) utilized in equations (4) - (7) for achieving theoretical values shown in Figure 6

| Name | $\omega$ (°) | $\Delta$ (Å) | $h_d$ (Å) | $d$ (Å) | $\sigma_{y0}$ (GPa) |
|---|---|---|---|---|---|
| Zigzag (Mo5\|7+S5\|7) | 17.45 | 6.00 | 11.43 | 5.93 | 13.0 |
| Armchair (Mo5\|7) | 21.98 | 3.34 | 8.37 | 1.63 | 18.3 |
| Armchair (S5\|7) | 21.98 | 3.23 | 8.36 | 1.65 | 22.1 |

**Figures and caption**

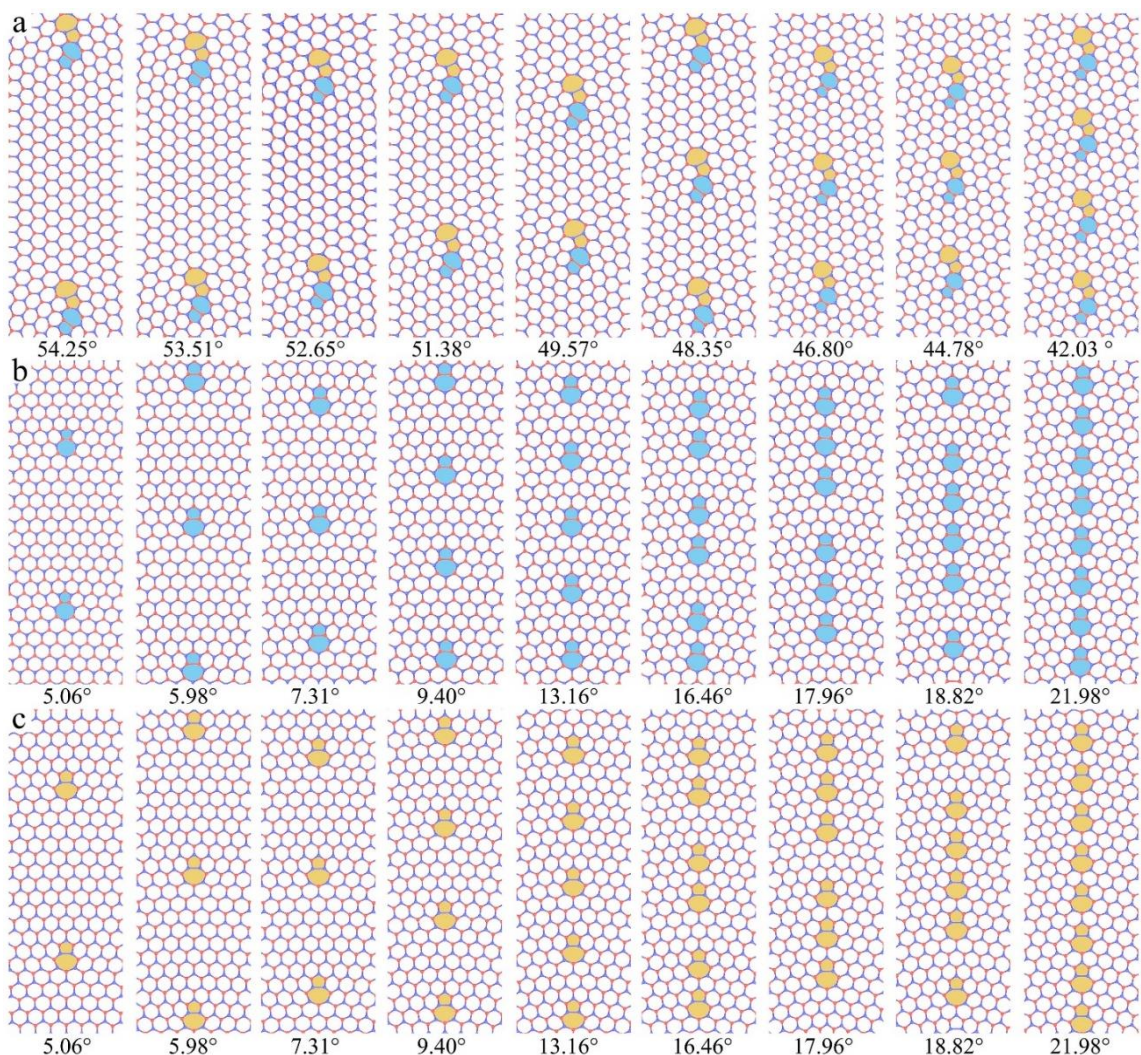

Figure 1 Grain boundary (GB) configuration of single-layer MoS$_2$ (SLMoS$_2$). (a) zigzag-tilt GBs with tilt angle varying from 42.03 to 54.25°, composed of M5|7+S5|7 dislocations. (b) and (c) Armchair-tilt GBs with tilt angle varying from 5.06 to 21.98°, in which they consist of Mo5|7 and S5|7 dislocations. For clarification, Mo5|7 and S5|7 dislocations in GBs are blue- and yellow-painted, respectively

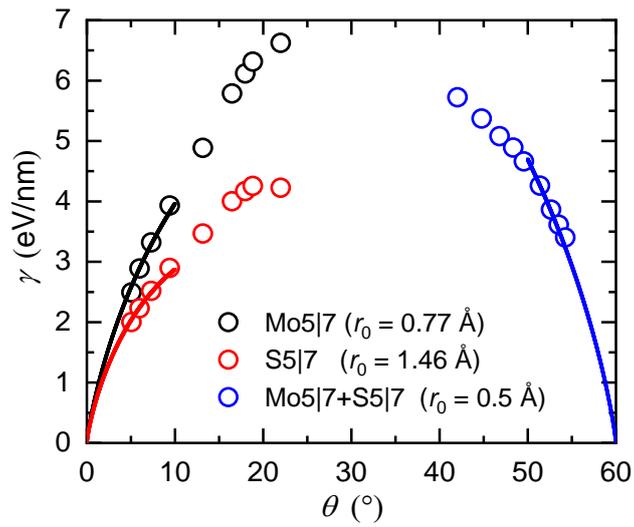

Figure 2 Energetics of grain boundaries (GBs) in single-layer MoS$_2$ (SLMoS$_2$). Variation in the energy of defective GBs with tilt angle. Black, red and blue solid circles indicate Mo5|7, S5|7 and Mo5|7+S5|7 dislocation-dominated GBs. The solid curves indicate the fittings from the Read-Shockley model.

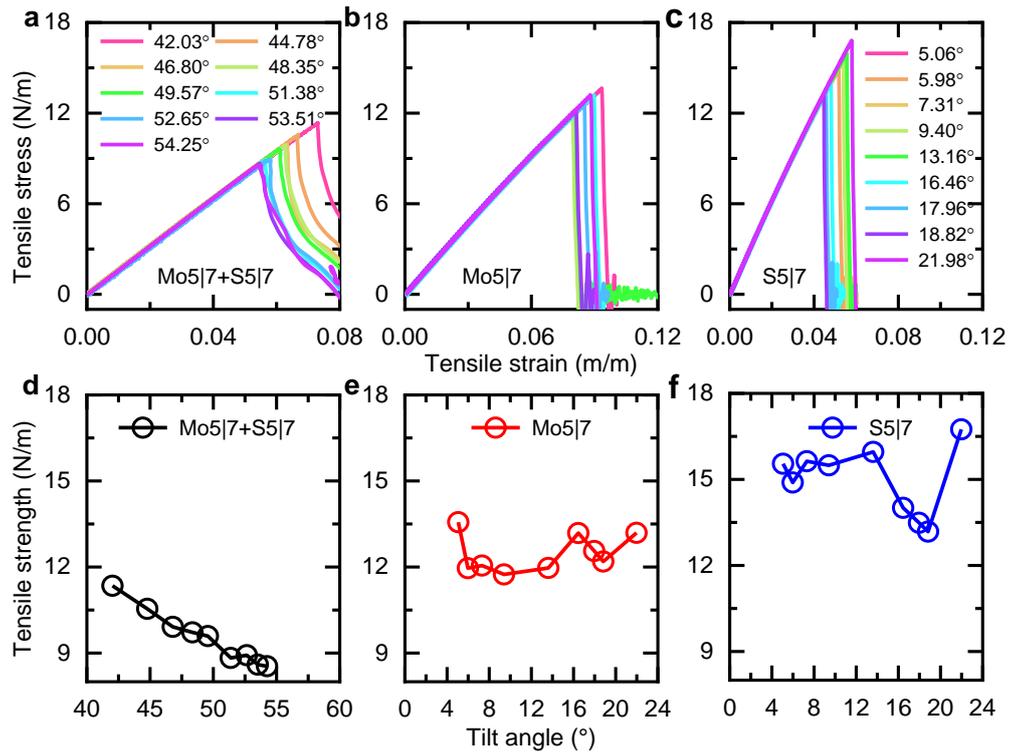

Figure 3 Mechanical properties of grain boundaries (GBs) in single-layer $MoS_2$ ($SLMoS_2$). (a) - (c) Tensile stress - strain curves of $SLMoS_2$ containing Mo5|7+S5|7, Mo5|7 and S5|7 dislocation-dominated GBs, respectively. (d) - (e) Variation in the tensile strength of Mo5|7+S5|7, Mo5|7 and S5|7 dislocation-dominated GBs with tilt angle, respectively.

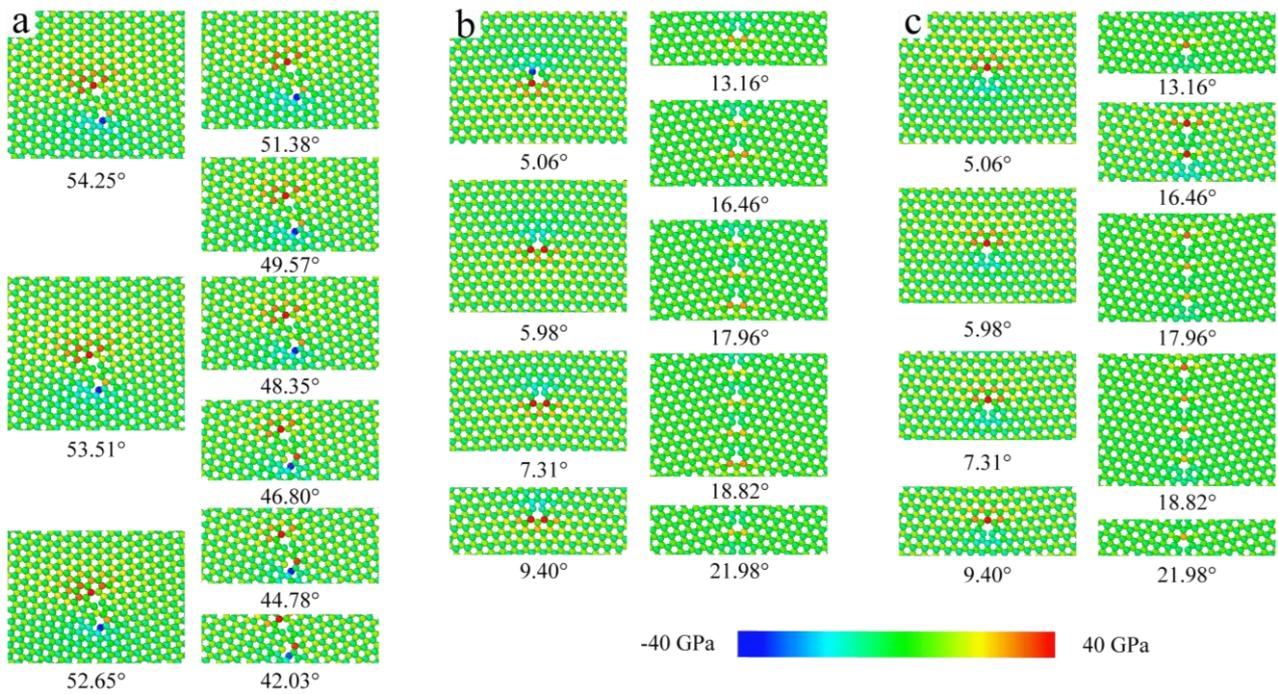

Figure 4 Pre-stress distribution at dislocation-dominated grain boundaries (GBs) in single-layer MoS2 (SLMoS$_2$). Contours of $\sigma_{xx}$ stress in the vicinity of (a) Mo5|7+S5|7, (b) Mo5|7 and (c) S5|7 dislocation-dominated SLMoS$_2$ GBs with different tilt angles at equilibrium state.

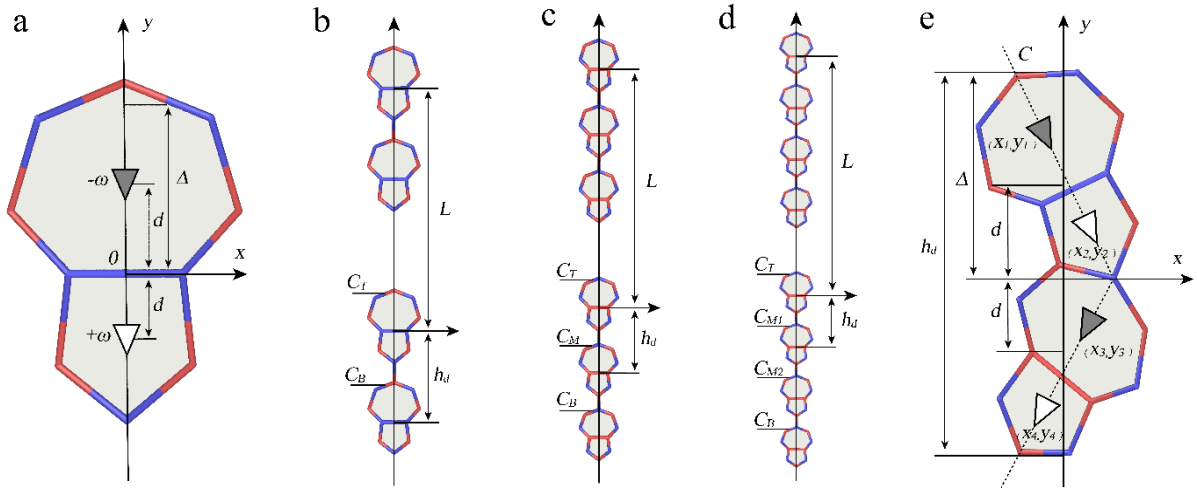

Figure 5 Schematic sketches of typical 5|7 dislocation in defective grain boundaries (GBs) of single-layer MoS$_2$ (SLMoS$_2$). (a) Isolated pentagon-heptagon 5|7 dislocation as a disclination dipole. (b) - (d) Configurations of 5|7 dislocations in armchair-tilt GBs with tilt angle of 16.46°, 17.96° and 18.82°, respectively, where $L = 2h_d + 3a$, $L = 3h_d + 3a$ and $L = 4h_d + 3a$. (e) Defective cluster made of two 5|7 dislocations in zigzag-tilt GBs, where $d$ is the half distance between the disclination centers. $\sigma_{y0}$ is the strength of the bond shared by pentagon and heptagon in the disclination dipole, which is not affected by other dipoles. $a$ is the bond distance of Mo-S.

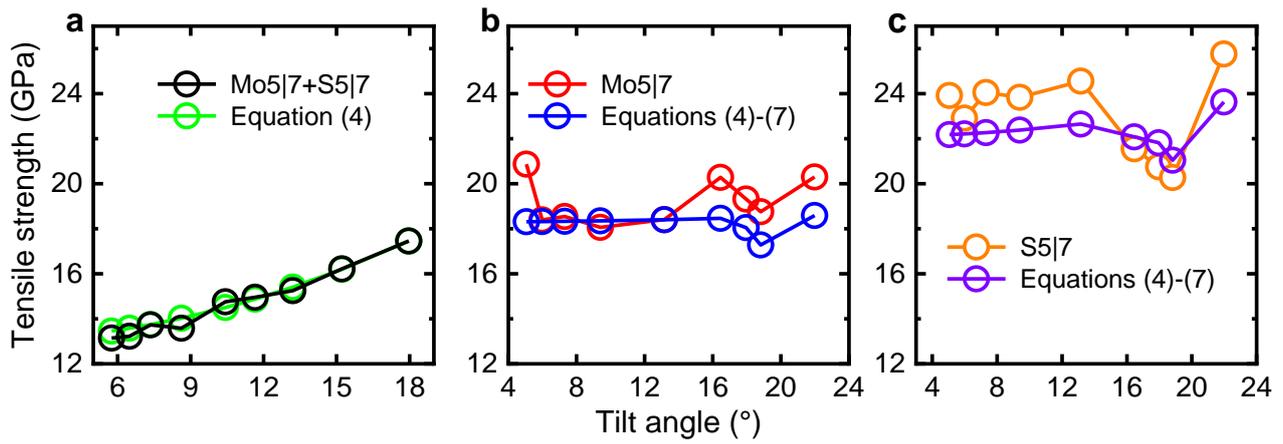

Figure 6 Comparison of MD predicted tensile strength and theoretical values of grain boundaries (GBs). (a) - (c) Variation in the tensile strength with tilt angle for Mo5|7+S5|7, Mo5|7 and S5|7 dislocation-dominated GBs. For Mo5|7+S5|7 dislocation-dominated GBs, the theoretical predictions are from Equation (4), whereas for Mo5|7 and S5|7 dislocation-dominated ones, they are from Equations (4)-(7).

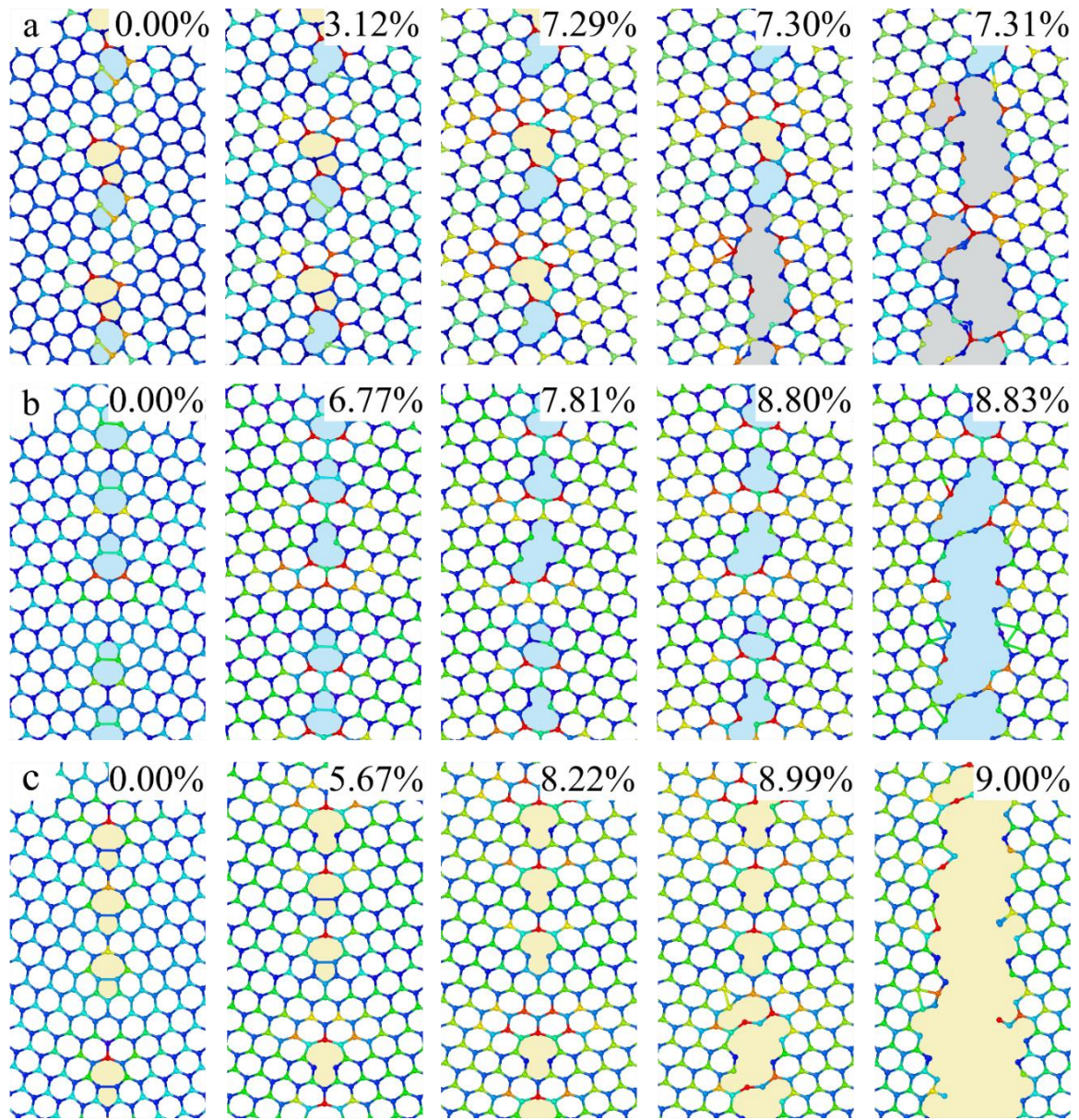

Figure 7 Snapshots of fracture during MoS$_2$ stretching. (a) zigzag GB with θ = 42.03°, (b) armchair Mo-rich GB with θ = 17.96°, (c) armchair S-rich GB with θ = 17.96°.

TOC figure

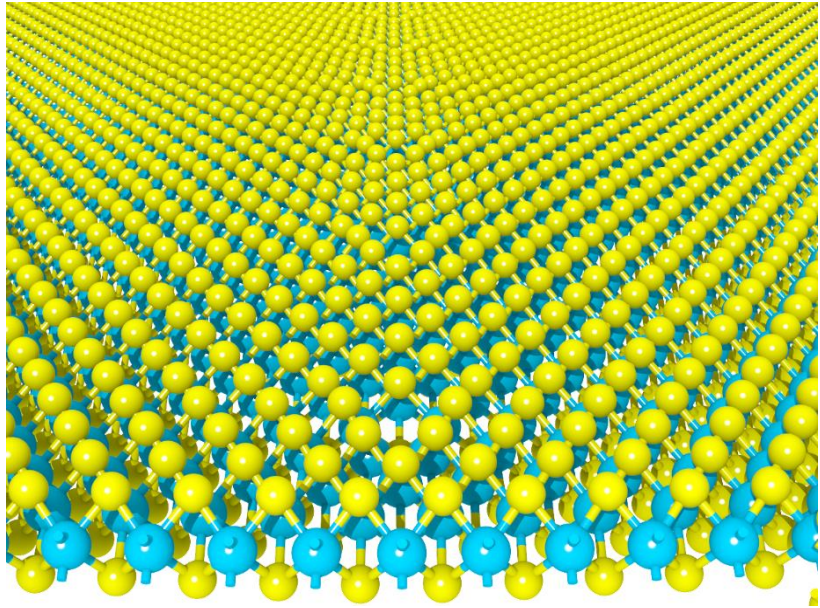